\documentclass[
reprint,
 amsmath,amssymb,
 aps,
 superscriptaddress,
 pra
]{revtex4-2}

\usepackage[utf8]{inputenc}
\usepackage{mathtools}
\usepackage[dvipsnames]{xcolor}
\usepackage{graphicx}
\usepackage{verbatim}
\usepackage[hidelinks]{hyperref}
\usepackage{appendix}

\usepackage[ruled]{algorithm2e}
\usepackage{newfloat,algcompatible}
\usepackage[noend]{algpseudocode}
\usepackage{braket}
\usepackage{bm}
\usepackage{cancel}

\usepackage{nicematrix}
\usepackage[normalem]{ulem}
\usepackage{dcolumn}

\newcommand{\Np}{{N_{\rm ph}}}
\newcommand{\methods}{{App. }}
\renewcommand{\Pr}{{\rm Pr}}

\begin{document}
\preprint{APS/123-QED}

\title{Multiphoton quantum simulation of the generalized Hopfield memory model}

\author{Gennaro Zanfardino}
\affiliation{Dipartimento di Ingegneria Industriale, Università degli Studi di Salerno, Via Giovanni Paolo II, 132, 84084 Fisciano (SA), Italy}
\affiliation{Institute of Nanotechnology of the National Research Council of Italy, CNR-NANOTEC, Rome Unit, Piazzale A. Moro 5, I-00185, Rome, Italy }
\affiliation{INFN, Sezione di Napoli, Gruppo Collegato di Salerno, Italy}

\author{Stefano Paesani}
\affiliation{NNF Quantum Computing Programme, Niels Bohr Institute, University of Copenhagen, Blegdamsvej 17, DK-2100 Copenhagen Ø, Denmark}
\affiliation{Center for Hybrid Quantum Networks (Hy-Q), The Niels Bohr Institute, University~of~Copenhagen,  DK-2100  Copenhagen~{\O}, Denmark}

\author{Luca Leuzzi}
\affiliation{Institute of Nanotechnology of the National Research Council of Italy, CNR-NANOTEC, Rome Unit, Piazzale A. Moro 5, I-00185, Rome, Italy }
\affiliation{Department of Physics, Sapienza University of Rome, Piazzale A. Moro 5, I-00185, Rome, Italy}

\author{Raffaele Santagati}
\affiliation{Quantum Lab, Boehringer Ingelheim, 55218 Ingelheim am Rhein, Germany}

\author{Fabio Sciarrino}
\affiliation{Department of Physics, Sapienza University of Rome, Piazzale A. Moro 5, I-00185, Rome, Italy}

\author{Fabrizio Illuminati}
\affiliation{Institute of Nanotechnology of the National Research Council of Italy, CNR-NANOTEC, Lecce Central Unit, c/o Campus Ecotekne, Via Monteroni, 73100 Lecce, Italy}
\affiliation{Dipartimento di Ingegneria Industriale, Università degli Studi di Salerno, Via Giovanni Paolo II, 132, 84084 Fisciano (SA), Italy}

\author{Giancarlo Ruocco}
\affiliation{Center for Life Nano- \& Neuro-Science, Italian Institute of Technology, IIT, Rome, Italy}
\affiliation{Department of Physics, Sapienza University of Rome, Piazzale A. Moro 5, I-00185, Rome, Italy}

\author{Marco Leonetti}
\email{marco.leonetti@cnr.it}
\affiliation{Institute of Nanotechnology of the National Research Council of Italy, CNR-NANOTEC, Rome Unit, Piazzale A. Moro 5, I-00185, Rome, Italy }
\affiliation{Center for Life Nano- \& Neuro-Science, Italian Institute of Technology, IIT, Rome, Italy}


\date{\today}

\begin{abstract}
In the present work, we introduce, develop, and investigate a connection between multiphoton quantum interference, a core element of emerging photonic quantum technologies, and Hopfield-like Hamiltonians of classical neural networks, the paradigmatic models for associative memory and machine learning in systems of artificial intelligence.
Specifically, we show that combining a system composed of $\Np$ indistinguishable photons in superposition over $M$ field modes, a controlled array of $M$ binary phase-shifters, and a linear-optical interferometer, yields output photon statistics described by means of a $p$-body Hopfield Hamiltonian of $M$ Ising-like neurons $\pm 1$, with $p = 2\Np$.
We investigate in detail the generalized $4$-body Hopfield model obtained through this procedure and show that it realizes a transition from a memory retrieval to a memory black-out regime, i.e. a spin-glass phase, as the amount of stored memory increases.
The mapping enables novel routes to the realization and investigation of disordered and complex classical systems via efficient photonic quantum simulators, as well as the description of aspects of structured photonic systems in terms of classical spin Hamiltonians.
\end{abstract}

\maketitle


\section{Introduction}

In recent years, the study of bosonic quantum interference with photons coherently evolving through linear-optical networks is of central importance in the current development of photonic quantum technologies.
This phenomenon implies a wide range of applications, going from architectures for fault-tolerant photonic quantum computing~\cite{knill2001scheme, GimenoSegovia2015, bartolucci2023fusion} to potential near-term quantum computational advantages with specialized algorithms~\cite{aaronson2013bosonsampling} and high-precision sensing~\cite{pirandola2018advances}.
Recent experimental advances have enabled the generation, interference, and detection of systems with over $100$ photons, achieving quantum computations at scales competitive with conventional supercomputers~\cite{zhong2020quantum, madsen2022quantum, aspuru2012photonic}.
In the present work, we establish and investigate the relation between multiphoton quantum interference in linear optics and generalized $p$-body Hopfield models of statistical physics \cite{Peretto1986,Baldi1987,Gardner1987,Abbott1988,Horn1988,krotov2016dense,ramsauer2021}.

The Hopfield model (HM) \cite{hopfield1982neural} is a fundamental tool in the study and characterization of memory storage and retrieval in biological neural networks \cite{Amit1989, muller1995neural,hertz2018introduction}, as well as in the investigation of machine learning problems \cite{Hinton2006,Hinton2007}. The HM is a full feedback neural network providing pattern recognition, associative memory functions, error correction, and other capabilities \cite{Amit1985, amit1987statistical}.
Generalized Hopfield Models (gHMs) with multi-synaptic interaction in the structure of deep neural networks are employed in studying information storage and prototype learning \cite{krotov2016dense, ramsauer2021}. Understanding the different phase regimes and collective behaviors associated with HMs and gHMs is thus a problem at the cutting edge in the study of artificial intelligence.

\begin{figure*}[ht!]
    \centering
    \includegraphics[width=.9\textwidth]{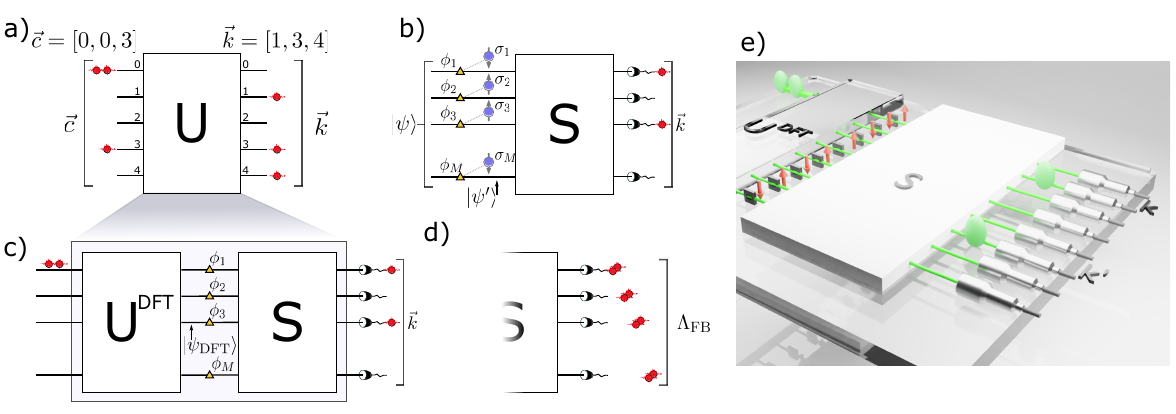}
    \caption{
    a) Standard description of linear optical transformation from an input configuration $\vec{c}$ to an output configuration $\vec{k}$ through a linear scattering matrix $U$.
    An example here is depicted for $\Np=3$ photons and $M=5$ modes.
    b) Schematic of the mapping of an $p$HM to a photonic system composed of an input state $\ket{\psi}$, a set of $M$ phase-shifters which can have binary values $\phi_i \in \{0, \pi \}$ and map to the $M$ spin states $\sigma_i = \text{exp}(i \phi_i)=\pm 1$, and a scattering matrix $S$.
    c) Full schematic  including the initial Discrete Fourier Transform $U^{\text{DFT}}$ for generating the near-uniform input state $\ket{\psi_\text{DFT}}$.
    d) Simplified scheme with the set of output states $\Lambda_{\text{FB}}$ given by fully-bunched configurations with all photons exiting in the same output mode. Panel e) reports the ``on-chip'' experimental scheme, correspondent to c).}
\label{fig:lin_optics}
\end{figure*}

Here we show how photonic architectures with a structure (cf. Fig.~\ref{fig:lin_optics}a-d) composed of i) $N_\text{ph}$ photons in an initial superposition state over $M$ optical field modes, ii) a layer of $M$ phase shifters (one on each mode), and iii) a linear-optical interferometer implementing a general scattering matrix over the $M$ modes, display photon statistics in the output modes that can be described through the Hamiltonian of a $p=2 N_\text{ph}$-body Hopfield model of $M$ neurons.

The $p=2$ HM has been first optically simulated with incoherent light \cite{farhat1985optical}, and successively with coherent light and a digital micromirror device architectures \cite{leonetti2021optical}, showing speed advantages for the all-optical simulations of large-spin systems. Furthermore, Hopfield-inspired memory storage and computation has been recently developed \cite{leonetti2024photonic}.
In these cases, as in the original HM \cite{hopfield1982neural}, the number $P$ of memories that can be retrieved is a fraction of the number of neurons in the network, $\alpha={P}/M < 1$ and cannot grow more than linearly with $M$.

Instead, considering energy functions
with interactions of order
$p>2$, it is possible to realize neural networks capable of  retrieving a much larger number of stored memory patterns
\cite{krotov2016dense,Gardner1987,Horn1988}.
 One drawback is that simulating the dynamics of fully connected networks with multi-neuronal synaptic couplings (i.e., in which $p$ neurons interact simultaneously) requires a time that scales super-extensively with the number of neurons. Generically, the computation time of the energy gradient in a dynamical iteration grows like $M^{p-1}$, i.e., as the average connectivity per neuron.  However, the analog photonic computation would drastically reduce this time.   To generalize to the multi-synaptic large storage case, though, requires a further stratagem in the photonic system proposed for the energy measurement: the use of entangled photons and interferometry.

The connection between linear-optical quantum systems and models from statistical physics opens ways  for investigating classical $p$HM with quantum photonic systems, potentially enabling efficient calculations for large $p$ and $M$.
We illustrate this through the simulation of photonic systems implementing a $4$HM, showing evidence for the characteristic transition from retrieval to spin-glass phase emerging from the photons' statistics.

\section{Photonic mapping of the generalized Hopfield model}
The type of physical system we consider for the mapping is a linear optical quantum processor, composed of single photons as inputs, a linear interferometer, which can universally be implemented with only phase-shifters and beam-splitters~\cite{reck1994, clements2016}, and photo-detectors.
Such a platform is, in principle, compatible with scalable quantum computers~\cite{knill2001scheme, GimenoSegovia2015, bartolucci2023fusion} as well as near-term quantum simulators~\cite{sparrow2018}, which is the type of application considered here. The advantages of photonic implementations include lighting fast transformations, inherent parallel operations, and low consumption, which are, in general, valuable properties for future generations of computing hardware~\cite{leonetti2024photonic,leonetti2021optical,fan2023photonic, pierangeli2019large}.
Here, we propose to use it as a new platform for optical computing in which we employ the inherent quantum properties of bosonic particles to map the classical $p$HM (see \methods  \ref{MODEL}).

The proposed mapping to a photonic system is performed by feeding a prepared state of $\Np$ single photons over $M$ optical modes to an interferometric network and enables the calculation of the energy and simulate the dynamics of a corresponding statistical physical $2\Np$HM.
The mapping stems from the equation governing the $\Np$-photon current after a linear transformation, such as the transmission through a random medium or a multi-mode fiber, characterized by a scattering matrix.

A single photon $j$ can be in $M$ possible modes, labeled by $c_j=0,\ldots, M-1$.
The Fock state of the $\Np$ photons is, then, represented by the vector $\vec{c}=\{c_1,\ldots,c_{\Np}\}$.
The set of all possible configurations of $\Np$ photons over $M$ modes is denoted by the set $C_M$, whose cardinality is
$|C_M|= \binom{\Np + M -1 }{\Np} \simeq M^{\Np}/\Np!$ for large $M$.

Considering a general photonic system as in Fig.~\ref{fig:lin_optics}a, the scattering amplitude of an input photon configuration $\vec{c}$ to an output configuration $\vec{k}$, when evolved through the transformation $\mathcal{U}$ described by the scattering matrix $U$, is given by~\cite{scheel2004permanents}:
\begin{equation}
        \bra{\vec{k}} \mathcal{U}(\ket{\vec{c}}) = \frac{1}{\sqrt{\mu(\vec{c}) \mu(\vec{k})}} \text{Perm} \left(U_{\vec{k}|\vec{c}}\right),
    \label{eq:perm_ampls}
\end{equation}
where $\text{Perm}(U_{\vec k|\vec c})$ is the permanent function of the $\Np\times \Np$ submatrix $U_{\vec k|\vec c}$ obtained by its rows (columns) given the elements in $\vec{k}$ ($\vec{c}$), cf. \methods  \ref{Sec:DFT}.
The factor $\mu(\vec c)\equiv \prod_{j=0}^{M-1} n_{j}! $, where $n_{j}=0,1,\ldots, \Np$, is the multiplicity of the mode $j$ in the $\Np$-photon configuration. 
For example, considering configurations with $\Np=3$, if all photons are in different modes then $\mu(\vec c)=1$, whereas if $c_1=c_2\neq c_3$ then $\mu(\vec c)=2$ (see Fig.~\ref{fig:lin_optics}a),  and if $c_1=c_2=c_3$ then $\mu(\vec c)=3$.
The transformation $\mathcal{U}$ is a linear function of the input configuration as well as the scattering matrix $U$ and returns the output state given, in general,  by a superposition of various output configurations $\vec k$.

 If the system is prepared in an input configuration $\vec{c}$, the probability that a configuration $\vec{k}$ is detected at the output of the transformation is thus given by
\begin{equation}
    \Pr_{\mathcal U}(\vec{k}|\vec{c}) = \left|\bra{\vec{k}} \mathcal{U}(\ket{\vec{c}})\right|^2 = \frac{1}{\mu(\vec{c}) \mu(\vec{k})} \left|\text{Perm} \left(U_{\vec{k}|\vec{c}}\right)\right|^2.
    \label{eq:perm_probs}
\end{equation}

To map the $p$HM,  we introduce a set of controllable elements composing the neuronal state, as shown in Fig.~\ref{fig:lin_optics}b.
The first is the initial state in which the photons are prepared. We consider a general superposition state  $\ket{\psi}= \sum_{\vec{c}\in C_M} a_{\vec{c}} \ket{\vec{c}}$ of $\Np$ indistinguishable photons over $M$ modes with amplitudes $a_{\vec{c}}$. State normalization gives $\sum_{\vec{c}\in C_M} |a_{\vec{c}} |^2 = 1$.
The state, then, undergoes a set of single-mode linear phase-shifters,  providing the opportunity for individual phase control, enabling to set an additive phase delay $\phi_{c}$ for each mode $c$.
For our mapping, the phases can take value $0$ or $\pi$ and transform the state as $\ket{\psi'}= \sum_{\vec{c}\in C_M} a'_{\vec{c}} \ket{\vec{c}}$ with new amplitudes $a'_{\vec{c}}$ given by
\begin{equation}
    a'_{\vec{c}}  =  a_{\vec{c}} e^{\imath \sum_{j=1}^{\Np} \phi_{c_j}}   = a_{\vec{c}}\prod_{j=1}^{\Np} e^{\imath \phi_{c_j} }= a_{\vec{c}}~\prod_{j=1}^{\Np} \sigma_{c_j} ,
    \label{def:spins-amplitude}
\end{equation}
\noindent
where $\sigma_c = e^{\imath \phi_c}=\pm 1$. These phases will correspond to {\em Ising spin} variables playing the role of the {\em neuronal activity} of the {\em neuron} $c$.

The system is then evolved through
a scattering medium with associated matrix $S$ and transformation $\mathcal{S}$.
The scattering amplitude for a final state $\ket{\vec{k}}$ from the input state $|\psi\rangle $ can be readily calculated using the linearity of the transformation $\mathcal S$ and Eqs. (\ref{eq:perm_probs},\ref{def:spins-amplitude}) as
\begin{align}
\label{eq:amplitude-psi}
    \bra{\vec{k}} \mathcal{S}(\ket{\psi'}) &=
    \sum_{\vec{c}\in C_M}  \frac{ a_{\vec{c}} }{\sqrt{\mu(\vec{c}) \mu(\vec{k})}} \text{Perm} \left(S_{\vec{k}|\vec{c}}\right)\prod_{j=1}^{\Np}\sigma_{c_j}.
\end{align}
Finally, we consider the placement of photo-detectors on the output modes of the interferometer (see Fig.~\ref{fig:lin_optics}b), allowing us to measure the output probability for each configuration $\vec k$  in a subset $\Lambda \subseteq C_M$ of all possible output configurations.
Denoting as $\vec x$ and $\vec y$ the input photon configurations in Eq. (\ref{eq:amplitude-psi}),
the joint occurrence probability of detecting photons in any configuration $\vec k\in \Lambda$ is simply given by the sum of single scattering probabilities for each $\vec k$, cf. Eq. (\ref{eq:perm_probs}):
\begin{align}
\label{output_prob}
    \Pr(\Lambda |{\bm\sigma} ) = \sum_{\vec{k}\in \Lambda}\Pr_{\mathcal S} (\vec{k} |{\bm\sigma} ) =
     \sum_{\vec{x},\vec{y} \in C_M} J_{\Lambda}(\vec{x},\vec{y})\prod_{j=1}^{\Np}\sigma_{x_j}\sigma_{y_j} ,
\end{align}
where, by $\bm \sigma=\{\sigma_1,\ldots, \sigma_M\}$ we denote the set of all possible configurations of $0,\pi$ phase modulations for all  $M$ photonic modes in $\vec x$ and $\vec y$. The coupling tensor $J_\Lambda$ is given by the sum of Hebb-like \cite{Hebb1949,hopfield1982neural} terms
\begin{align}
    J_{\Lambda}(\vec{x}, \vec{y}) = \frac{a_{\vec{x}} a^*_{\vec{y}}}{\sqrt{\mu(\vec x) \mu(\vec y)}}
    \sum_{\vec{k}\in \Lambda}    \frac{ \text{Perm} \left(S_{\vec{k}|\vec{x}}\right) \text{Perm}^*\left(S_{\vec{k}|\vec{y}}\right) }{\mu(\vec{k})}.
    \label{eq:J_ensemble}
\end{align}
Eq. (\ref{output_prob}) can be related to a  $p$HM Hamiltonian, with $p=2N_{ph}$, as $H[\bm{\sigma}| \Lambda ]=-M \Pr( \Lambda |\bm{\sigma})$,
where $\Lambda$ plays the role of the set of memory patterns planted into the network.
The last term in Eq. (\ref{output_prob}) is a random walk of $|C_M|^2$ steps in the space of input states. As the number $M$ of photonic modes increases, in order for $\Pr(\Lambda|\vec{\bm \sigma})$ to remain of $O(1)$,  as it suits to a probability, the random coefficients $J_\Lambda$ must have a mean square displacement scaling for large $M$ like  $|J_\Lambda|\sim 1/|C_M|\sim 1/M^\Np$, as it might also be deduced from the scaling of the scattering matrix elements, $|S_{k_i x_j}|\sim 1/\sqrt{M}$ in Eq. (\ref{eq:J_ensemble}).

In the $2$ photons case, Eq. (\ref{eq:J_ensemble}) becomes a $4$-spin interaction, scaling as $1/M^2$:
\begin{widetext}
\begin{align}
    J_{\Lambda}(x_1, x_2, y_1, y_2) = \frac{ a_{\vec{x}} a^*_{\vec{y}} }{\sqrt{\mu(\vec{x}) \mu(\vec{y})}}
    \sum_{[k_1, k_2] \in \Lambda}
    \frac{
    \left[S_{k_1,x_1}S_{k_2,x_2} + S_{k_1,x_2}S_{k_2,x_1} \right]
    \left[S_{k_1,y_1}S_{k_2,y_2} + S_{k_1,y_2}S_{k_2,y_1} \right]^*}{\mu(\vec{k})}.
      \label{eq:J_ensemble_2_phot}
\end{align}
\end{widetext}
Note that, if the set of detectable output configurations $\Lambda$ is composed of a single configuration $\vec{k}_0$, then the
$p=2\Np$-neuron Hamiltonian (\ref{output_prob}) is factorized into  two $\Np$-body Hamiltonians   \cite{leonetti2021optical}.
%
On the other hand,
if $\Lambda$ encompasses all possible $P_{\rm all}\equiv \binom{\Np+M-1}{\Np}$ configurations of $\Np$
indistinguishable photons over $M$ modes, i.e. $\Lambda = C_M$,
we will simply have $\Pr(C_M |{\bm \sigma})=1$, implying a trivial  dependence on $\bm\sigma$.
Therefore, a choice of $P \equiv |\Lambda| < P_{\rm all}\sim M^{\Np}$ as $M\gg 1$ will be required to ensure the mapping of a non-trivial spin Hamiltonian.

Another important point to note from Eqs. (\ref{eq:J_ensemble},\ref{eq:J_ensemble_2_phot}) is that the coefficients $a_{\vec{x}}$ factor out in the coupling function.
This means, for example, that in order to have a fully connected neural network, the amplitudes $a_{\vec{x}}$ of each configuration $\vec{x}$ present in superposition in the input state need to be non-zero.
We describe a possible scheme for generating such states in the next section.

\section{Preparing uniform input states via a Discrete Fourier Transform}

In order to have a fully connected network
it is useful to have all coefficients $a_{\vec{x}}$ in the input superposition as uniform as possible.
A practical scheme to achieve this regime is shown in Fig.~\ref{fig:lin_optics}c, while its correspondent ``on chip'' 3d rendering  in Fig.~\ref{fig:lin_optics}e.
We consider the preparation of $\ket{\psi}$ by starting with all the $\Np$ photons in a single input mode of the interferometer, e.g., the $0$-th one, and transmitting them through a Discrete Fourier Transform (DFT) interferomenter~\cite{Crespi2016, wang2023deterministic}, implemented by the unitary transformation
$U^{\text{DFT}}_{jq} = e^{2\pi\imath \frac{ jq}{M}}/\sqrt{M}$.
In free space optics, this configuration corresponds to a source from a single-mode fiber coupled with a collimation lens  (see \methods  \ref{EXPERIMENT}). Applying this transformation to the input Fock state with all photons in, e.g., mode zero, prepares a superposition state with amplitude $a_{\vec{x}}= \sqrt{\Np! /  (M^{\Np}\mu(\vec{x}))}$ for any configuration $\vec{x}\in C_M$ (see \methods  \ref{Sec:DFT}). These amplitudes are all non-zero and uniform up to the multiplicity factor $1/\sqrt{\mu(\vec{x})}$, thus ensuring a complex structure in the mapped synaptic couplings $J$.

\begin{figure}[ht!]
    \centering
    \includegraphics[width=8cm]{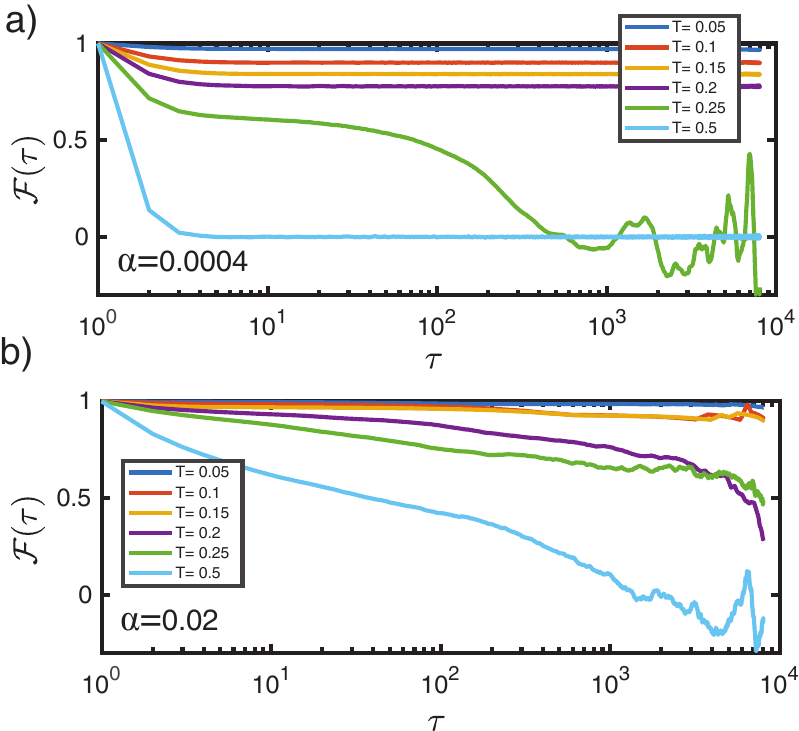}
    \caption{ Panels a) and b) report $\mathcal{F(\tau)}$ for various temperatures (see legend), for values of the storage size ration $\alpha=0.0004$ (retrieval regime) and  $\alpha=0.02$ (spin-glass phase), respectively.
     Panel c) compares the Metropolis standard deviation $\sigma_T=\sigma_{\rm Metropolis}$ with the experimental standard deviation  $\sigma_{Exp}$ resulting from a limited number of photon pairs per iteration step. $\sigma_{Exp}$ for various "Experimental time per iteration"  (ETI) is reported, assuming a source of photon pairs capable of generating 20 million photon pairs per second. }
     \label{fig:FselfError}
\end{figure}

\begin{figure*}[ht!]
    \centering
    \includegraphics[width=15.5 cm]{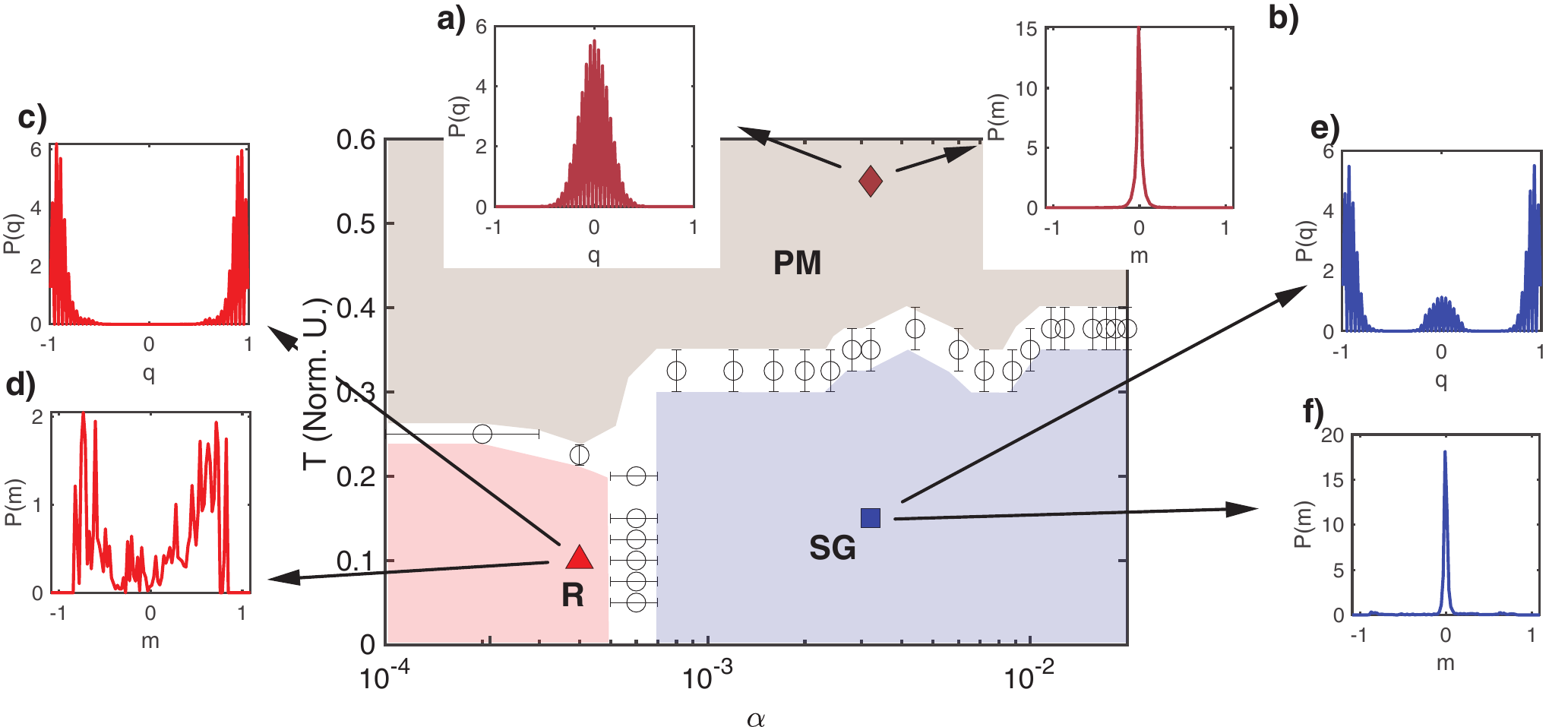}
    \caption{Phase Diagram of the $4$-Hopfield model  realized with a quantum interferometer of $2$-photons on $M=50$ modes. The insets report $P(q)$ and $P(m)$ for the paramagnetic phase (a, b, $\alpha=0.0032$ ; $T= 0.55$), retrieval phase (c, d $\alpha=0.0004$ ; $T= 0.1$), and spin glass phase (e, f $\alpha=0.0032$ ; $T= 0.15$).} 
    \label{fig: PhDiag}
\end{figure*}

\section{Metropolis photonic simulation of the classical spin system}

The remainder of the manuscript will focus on the  $\Np=2$ case, mapping into a $4$HM, cf. Eq.(\ref{eq:J_ensemble_2_phot}).
This equivalence can be exploited to extract the energy of a $4$HM  by the analog measurement of the occurrence probability Eq.(\ref{output_prob}).
The simulations reported here have been realized with the fully bunched input configuration, shown in Fig. \ref{fig:lin_optics}d and detailed in \methods  \ref {FULLYBUNCHED}.

In a single experimental iteration $n-1\to n$, one of the modes' phases is flipped ($0 \rightarrow \pi$ or $\pi \rightarrow 0$), and then, the new two-photon current is measured. The change is straightforwardly accepted in the simulated dynamics if $\mathcal H$ decreases, meaning $\Pr (\Lambda |\vec{\sigma} )$ increases.
If we introduce synaptic noise, though, in the form of a statistical mechanical {\em temperature}  $T$, the change can also be accepted if the energy cost momentarily {\em increases},
according to the Markov Chain Monte Carlo Metropolis algorithm \cite{Metropolis53}.
In our notation, the Metropolis prescription for a phase change, given $\Delta \mathcal H \equiv \mathcal H[{\bm \sigma}(n)]-\mathcal H[{\bm \sigma}(n-1)]$, reads
\begin{eqnarray}
 \mbox{if } \Delta \mathcal H < 0 && \quad \mbox{change accepted with prob. $1\phantom{^{-\Delta\mathcal H/T}}$} \qquad \\
  \mbox{if } \Delta \mathcal H > 0  && \quad \mbox{change accepted with prob. $~e^{-\Delta\mathcal H/T}$}. \qquad
  \label{metropolis}
\end{eqnarray}
At each Monte Carlo step $t$ (sequence of $M$ single phase shift iterations),  the ${\bm\sigma}(t)$ configuration is recorded. Different time configurations are employed to compute the self-correlation function $F_{self}(\tau)=\braket{\bm\sigma(t) \cdot \bm\sigma(t+\tau)}$ (see \methods  \ref{SELFCORR}).

Results for $F_{self}(\tau)$ for various temperatures and two different values of the storage size ratio  $\alpha\equiv
P/M^{\Np}= P/M^{2}$  are reported in Fig. \ref{fig:FselfError} for $M=50$.
The parameter $\alpha$ is the ratio between the number of output scattering patterns of the $2$-photon current, $P=|\Lambda|$,
and the large $M$ scaling of the cardinality of $|C_M|$ (all  $N_{\rm ph}$-photon Fock states), apart from a constant factor.
i.e., $P_{\rm all}\sim M^{p/2}= M^2$ \footnote{This is a difference with respect to the standard generalized $p$-spin Hopfield models with $P$ stored memory patterns and $M$ neurons in which though with different prefactors in the model definitions \cite{Gardner1987, Abbott1988, Horn1988}, the scaling is always $\alpha \sim P/M^{p-1}$.}.

For $\alpha=0.0004$, corresponding to a single $2$-photon output channel, in Panel \ref{fig:FselfError}a) we can see that $F_{\rm self}(\tau)$ tends to plateau if  $T$ is low enough, thus displaying the relaxation to the unique fixed point corresponding to memory retrieval. As the synaptic noise increases, at some critical $T$ the $\bm \sigma$ configuration decorrelates to zero (faster for higher $T$) and retrieval is eventually lost.
As $\alpha$  increases (Panel \ref{fig:FselfError}b)  $F_{self}(\tau)$ decreases to zero for gradually lower and lower temperatures, and the memory retrieval becomes progressively harder.
The onset of spurious excited states \cite{Amit1989},  furthermore, attracts the dynamics to attractors not corresponding to the scattering patterns embedded in the network through Eq. (\ref{eq:J_ensemble_2_phot}).
The latter are the global minima of the energy $\mathcal H[\bm\sigma|\Lambda]$ landscape, whereas the former are  excited states corresponding to local minima.
Their existence implies that a plateau of
$F_{\rm self}(\tau)$
may not correspond to retrieval if the initial configuration
$\bm\sigma (0)$ lies in the {\em basin of attraction} of a spurious state.
Eventually, when $\alpha$ is larger than a critical value the onset of a spin-glass phase due to the proliferation of spurious states leads to memory black-out  \cite{Amit1985}.

Panel \ref{fig:FselfError}c) reports the comparison between the  fluctuations induced by the nonzero synaptic noise ($T>0$),
and those induced by the measurement noise (see \methods  \ref{NOISE}).
In the panel, open circles represent the standard deviation $\sigma_T$ of the temperature-induced energy fluctuations versus temperature. Continuous lines, instead, are relative to the standard deviation $\sigma_{\rm exp}$ of the experimental noise.
The proposed analog simulation is, indeed, valid only when the optical experimental measurement enables to assess the value of $\Pr(\Lambda |{\bm\sigma}(t))$ with an error negligible with respect to the fluctuations induced by the simulation temperature: $\sigma_{\rm exp}<\sigma_T$.
By increasing the experimental ``exposure time'' (controlling the number of measured photon pairs given a source with a finite production of pairs per second), one can lower the $\sigma_{\exp}$, thus enlarging the available measurement window.

\section{Analysis of spin-glass phase transitions}

An interesting property of the HM is the emergence of spin-glass phase transitions as $\alpha$ increases. Here, we analyze the behavior of the spin systems mapped through our approach and show numerical evidence of such phase transition. In order to explore the full phase-space through numerical simulations on a classical computer, we consider a configuration of the photonic system, which is advantageous for classical simulations,  shown in Fig.~\ref{fig:lin_optics}d.
In particular, we consider the case where the largest set $\Lambda =\{(0,0), (1,1), \ldots, (M-1,M-1)\} \equiv \Lambda_{\text{FB}}$ is composed of all configurations where the photons are fully-bunched together in any of the output modes.
Indeed, in this case, the output probability for any values of $M$ and $\Np$, cf. Eqs. (\ref{output_prob},\ref{eq:J_ensemble}), for a $\Lambda \subseteq \Lambda_{\rm FB}$ reduces to 
(see \methods \ref{FULLYBUNCHED})

\begin{equation}
    \Pr(\Lambda|{ \bm \sigma}) = \frac{1}{M^\Np} \sum_{k\in [0,M-1]}
    \left| \sum_{j=0}^{M-1} S_{k,j} \sigma_j \right|^{2\Np},
    \label{eq:compiled_prob}
\end{equation}
\noindent  with indexes $k$ running on a subset of optical modes and $j$ running through all modes. Considering this model significantly speeds up classical simulation
while still maintaining the whole structure of the mapped HM. We employ it here to study its phase diagram in $4$HM case. According to Eq. (\ref{eq:compiled_prob}) we have access to a storage coefficient $\alpha
\leq 1/M$, e.g., $\alpha \leq 0.02$ for $M=50$.
The Monte-Carlo simulations are implemented using the Exchange Monte-Carlo (EMC) method~\cite{hukushima1996exchange} with simultaneous system dynamics at different temperatures running in parallel over multiple CPUs. All results are obtained considering a system size of $M=50$ spins, a total of $2 \times 10^5$ simulated Monte-Carlo steps for system thermalization and exchanges between system dynamics at different temperatures are attempted every $200$ Monte-Carlo steps. 
The scattering matrix $S$ is taken randomly from the Haar random distribution, and all EMC simulations are repeated for $20$ different disordered samples, each denoted by an apart generated $S$ matrix. For each scattering sample, a total of $36$ replicas are simulated for each value of $\alpha$ and $T$ to extract the order parameters.
 In particular, the two order parameters we consider are standard parameters for spin-glass and memory retrieval systems: the overlap between simulated replicas $q_{ab}$ and the {\em memory} overlap $\hat{m}_{\vec{k}}$ (see \methods  \ref{OVERLAP-M}, \ref{OVERLAP-Q} for their definitions).

The results are presented in Fig.~\ref{fig: PhDiag}. We identify three phases: the retrieval phase, the spin-glass phase and the paramagnetic phase.
The Retrieval (R) phase is characterized by  large values of the memory overlap $\hat m_{\vec k}$, cf.  Eqs. (\ref{permanent-overlap},\ref{memory-overlap}) in \methods  \ref{OVERLAP-M}  and describing the degree of similarity of the retrieved state $\bm\sigma$ to the stored memory represented by permanents of the scattering matrix.
In inset $d$ of Fig. \ref{fig: PhDiag}, the probability distribution of the $\hat m_{\vec k}$ is displayed for a single configuration of the output $2$-photon current.
It has two non-trivially high peaks for large $|m|$, indicating the (noisy) alignment with the only pattern stored in the $4$HM.

The Parisi spin-glass overlap distribution, cf. Eq. (\ref{Qab}) in \methods \ref{OVERLAP-Q}, is shown in panel $c$, displaying two symmetric peaks representing all replicas aligned or counter-aligned to the unique pattern. As the number of permanent patterns composing the synaptic couplings Eq. (\ref{eq:J_ensemble_2_phot}) increases, the system at low $T$ has a phase transition to the Spin Glass (SG), characterized by all memory overlaps being negligible, corresponding to a memory blackout, and a non-trivial $P(q)$ with both strong ($|q| \to 1$) and weak ($q\sim 0$) correlations between replicas. Increasing the temperature, a paramagnetic phase arises both from the retrieval and the spin-glass phase.
This phase is characterized by a complete independence of all spins/neurons due to the high temperature/high synaptic noise and manifested in $P(q)\sim P(m)$ strongly peaked around $0$.

We highlight that by increasing the number of degrees of freedom $M$, the retrieval phase should shrink towards the $\alpha=0$ axis. This comes about because the permanents of the scattering matrices, playing the role of generalized memory patterns in the Hopfield analogy, are continuous (complex) numbers, and the critical behavior will not be qualitatively different from the case of Gaussian memory patterns already studied by Amit, Gutfreund and Sompolinsky \cite{Amit1985,amit1985b}.

\section{Discussion and Outlook}

The scheme introduced in the present work allows us to relate a structured quantum photonic interferometer to a $p$-body Hopfield model of associative memory. We numerically investigated  the mapping  with $p=4$, characterizing the phase diagram of the system with respect to the storage size ratio $\alpha$ and the temperature $T$, including a retrieval phase in the low $\alpha$-low $T$ region and a spin-glass phase in the high $\alpha$-low $T$ region.

In the present work, we have focused on systems implemented on complete graphs, where all spins are fully connected through a random scattering matrix $S$. One interesting future research direction is to investigate how spin Hamiltonians with other specific network structures, e.g., with local interactions or sparse random graphs, can be realized by different photonic architectures. For example, $p$-body models with local-only interactions might be engineered by considering a scattering matrix $S$ that only mixes optical modes associated with neighboring spins.
Similarly, input states with non-uniform amplitudes, in Eqs. (\ref{eq:J_ensemble},\ref{eq:J_ensemble_2_phot}), could be considered in order to suppress interaction terms in the Hamiltonian and yield a network structure with, e.g., diluted and distance-dependent interactions.

Another critical research direction enabled by our mapping is the possibility that photonic quantum architectures might simulate $p$HM and other classical neural networks more efficiently than the current existing methods on conventional computers.
Photonic hardware can be scaled up to systems containing millions of optical modes (Digital Micromirror Devices, see \methods  \ref{EXPERIMENT}) corresponding to $p$HM's with millions of spins.
This would allow us to tackle the simulation of models with super-extensive memory storage capability \cite{Agliari2020,Theriault2024} for neuronal set sizes $M$ comparable to, and even larger than those that are used to simulate pairwise Hopfield-like models on classical computers~\cite{FACHECHI2019,Leuzzi2022,Agliari2023,Agliari2024}. We propose an experimental set-up for implementing a photonic system in the \methods  \ref{EXPERIMENT}. Developing such photonic quantum simulators could potentially allow to probe computational regimes for simulating classical spin systems at or beyond the frontier of what is currently possible with conventional classical methods.

\section*{Ackowledgements}
S.P. acknowledges funding from the Marie Sklodowska-Curie Fellowship project QSun (Grant No. 101063763), the VILLUM FONDEN research grants No.~VIL50326 and No.~VIL60743, and support from the NNF Quantum Computing Programme; L.L. acknowledges funding from the Italian Ministry of University and Research, call PRIN 2022, project ``Complexity, disorder and fluctuations'', grant code 2022LMHTET;
G.R. Aknowledges European Research Council through its Synergy grant program, project ASTRA (grant agreement No. 855923) and by European Innovation Council through its Pathfinder Open Program, project ivBM-4PAP (grant agreement No. 101098989);
F.I. acknowledges funding from the Italian Ministry of University and Research, call PRIN PNRR 2022, project ``Harnessing topological phases for quantum technologies'', code P202253RLY, CUP D53D23016250001, and PNRR-NQSTI project "ECoN: End-to-end long-distance entanglement in quantum networks", CUP J13C22000680006;
M.L. acknowledges funding from the Italian Ministry of University and Research, call PRIN 2022, Project ``TARDIS: TArgeting misfolded Retinal tau for early AlzheimeR’s DIsease diagnosiS'' (2022CFP7RF  CUP: B53D23018560006 M.L.). This study was conducted using the  HPC infrastructure DARIAH of the National Research Council of Italy, by the CNR-NANOTEC in Lecce, funded by the MUR PON  ``Ricerca e Innovazione 2014-2020'' program, project code PIR01-00022.

\bibliography{BIB}

\pagebreak
\clearpage
\appendix
\section{Methods}

\subsection{The \textit{p}-body Hopfield Model}
\label{MODEL}
In the original HM \cite{hopfield1982neural} two neurons $i$ and $j$  are connected in couples by {\em symmetric} synapsis $J_{ij}$. The coupling
 values are constructed   starting from a number $P$ of given neuronal patterns $\bm \xi^{\mu}=\{\xi_1^{\mu},\ldots,\xi_M^{(\mu)}\}$, with $\xi_i^{(\mu)}=\pm 1$, representing memories intentionally planted into the network,
according to the Hebb rule \cite{Hebb1949}
\begin{equation}
J_{ij}= \frac{1}{M}\sum_{\mu=1}^{\mathcal{P}} \xi_i^{(\mu)} \xi_j^{(\mu)}
\label{eq:Jij}
\end{equation}

The neuronal activity $\bm \sigma$ is represented by a set of $M$ Ising spins $\sigma_i=\pm 1$, i.e., active or passive neurons.
A Hamiltonian $\mathcal{H}$ can be devised requiring that the planted memory patterns are {\em attractors} of the $\bm \sigma$ dynamics, i.e.,
the minima of $\mathcal H[\bm\sigma]$ are realized by those $\bm\sigma=\bm\xi^{(\mu)}$.
Introducing the memory overlap between the neuronal state $\bm \sigma$ and the memory pattern $\bm \xi^{(\mu)}$,
\begin{equation}
    \label{magnetization-2}
    m_\mu \equiv \frac{1}{M}\bm\sigma\cdot {\bm\xi}^{(\mu)}=\frac{1}{M}\sum_{i=1}^{M} \sigma_i\xi_i^{(\mu)} \in [-1,1],
\end{equation}
this reads
\begin{equation}
\mathcal{H}[\bm \sigma] =  -\frac{M}{2}\sum_{\mu=1}^{\mathcal P} \left(m_\mu^2-\frac{1}{M}\right)
=
-\frac{1}{2}\sum^{1,M}_{i\neq j} J_{ij}\sigma_i\sigma_j
\label{eq:Standard HM}
\end{equation}

The Hopfield model has been generalized to multi-neuron synapsis,  connecting $p>2$ neurons $i_1,\ldots,i_p$, with $J_{i_1,\ldots,i_p}\propto \sum_\mu \prod_{\ell=1}^p \xi_{i_\ell}^{(\mu)} / M^{p-1}$. The  Hamiltonian is  \cite{Peretto1986,Gardner1987,Abbott1988,Baldi1987,Horn1988}
\begin{equation}
\mathcal{H}[\bm\sigma]= -M \sum_{\mu= 1 }^\mathcal{P}  \left| m_\mu \right |^p + O\left(1\right) =
- \sum_{i_1, \ldots , i_p}^{1,M} J_{i_1,\ldots ,i_p} \prod_{\ell=1}^p \sigma_{i_\ell}.
\label{eq:General nu-HM}
\end{equation}
where $J$'s with equal neuron indices can be included or not, depending on the model choice, all-different-indices synaptic interactions yield the larger memory capacity~\cite{Horn1988}.

Synaptic nonlinearity (i.e., $p>2$)
realizes a larger memory capacity than in pairwise models. Indeed, there can be $M^p/p!$ Hebb-like synapsis and the upper limit of storage capacity would scale like $P_{\rm max}\sim M^{p-1}$~\cite{Amit1989}.

In general, nonlinearity boosts the recognition and classification capabilities of a network enabling the ``universal approximation'' of any function~\cite{krotov2016dense}.
A $p$HM, i.e.,  with $p-1$ rank nonlinearity, shows an improved storage capacity and finer pattern distinction capability, with the potential to boost the  future generation neural networks \cite{ramsauer2021}.
In this work, we realize  $p$-spin Hopfield-like models with even $p$ by means of entangled photons scattering in random media.

\subsection{Discrete Fourier Transform and input state}
\label{Sec:DFT}
In order to map the outcome probability in a $2\Np$-Hopfield Hamiltonian with $M$ degrees of freedom (activity of $M$ neurons), we need to distribute $\Np$ photons over $M$ modes. To do so, we start with a state in which all the $\Np$ photons are all in the mode 0:
\begin{equation}
    \ket{\vec{0}} = \frac{\big( \hat{a}_{1}^{\dagger} \big)^{\Np}}{\sqrt{\Np!}} \ket{\mathbf{0}} = \ket{\Np \dots 0}
\end{equation}
and we apply a Discrete Fourier Transform (DFT) unitary transformation, whose matrix elements are defined as:
\begin{equation}
U_{kl}^{DFT} = \frac{1}{\sqrt{M}} e^{-2\pi i \frac{k l}{M}}
\end{equation}
where: $ ( k, l ) \in \{0, 1, \dots, N-1\}$ and  $M$ is the size of the matrix that coincides with the total number of modes available.
\begin{equation}
    \hat{a}_{i} \overset{U^{DFT}}{\longrightarrow} \sum_{j} U_{ij}^{DFT} \hat{a}_{j}
\end{equation}
and all the photons in the source state are in the mode $0$.
Representing by matrix $U_{\vec x|\vec k}$ the $\Np\times\Np$ submatrix

\begin{equation}
\left(\begin{array}{cccc}
U_{x_1,k_1} & U_{x_1,k_2} &   \ldots   & U_{x_1,k_{\Np}} \\
U_{x_2,k_1} & U_{x_2,k_2} &   \ldots   & U_{x_2,k_{\Np}}  \\
\vdots        & \vdots        &            &  \vdots  \\
U_{x_{\Np},k_1} & U_{x_{\Np},k_2} &\ldots & U_{x_{\Np},k_{\Np}}
\end{array}\right) \quad ,
\end{equation}
\noindent obtained taking the rows (columns) of $U$ that correspond to the elements of $\vec{x}$ ($\vec{k}$), the coefficients of the state after the application of the DFT can be computed as
\begin{align}
    a_{\vec{x}} &= \bra{\vec{x}}\mathcal{U}(\ket{\vec{0}}) \notag \\
    &= \frac{1}{\sqrt{\Np! \mu(\vec{x}})} \text{Perm}\left(U_{\vec{x}|\vec{0}}^{DFT}\right) \notag \\
    &= \frac{1}{\sqrt{\Np! \mu(\vec{x}})} \text{Perm}\left(
    \begin{bNiceArray}{c|c|c|c}
    U_{\vec{x}|0}^{DFT} & U_{\vec{x}|0}^{DFT} & \ldots & U_{\vec{x}|0}^{DFT}
    \end{bNiceArray}\right) \notag \\
    & = \frac{1}{\sqrt{\Np! \mu(\vec{x})}} \sum_{\sigma\in \pi(\vec{x})} \prod_{i\in \vec{x}} U_{i, 0} \notag \\
    & = \sqrt{\frac{\Np!}{ \mu(\vec{x})}} \prod_{i\in \vec{x}} U_{i, 0}^{DFT} \notag \\
    & = \sqrt{\frac{\Np!}{M^{\Np} \mu(\vec{x})}}.
\end{align}
Hence the outcome state is
\begin{equation}
    \ket{\Psi} = \sum_{\vec{x}}a_{\vec{x}} \ket{\vec{x}}
\end{equation}
which is the desired result. At this stage, we apply the controlled-phase shift and the scattering medium $S$ to obtain the output probability given in Eq.~\eqref{output_prob}.


\subsection{Derivation of   $\Pr_{\mathcal S}(\vec{k}|\vec{c})$ }
\label{PS_DERIVATION}
The first derivation of the form of $\Pr_{\mathcal S}(\vec{k}|\vec{c})$ in the context of linear optics was obtained in Ref.~\cite{scheel2004permanents}. We describe it here for completeness. The evolution of (indistinguishable) photons through a scattering medium is described by a unitary matrix \( S \), under which the creation operators transform as
\begin{equation}
\label{Scattering_trans}
    \hat{a}_{i}^{\dagger} \to \sum_{j=1}^{M} S_{ji} \hat{a}_{j}^{\dagger}.
\end{equation}
The generic input state of $\Np$ photons distributed over $M$ modes is
\begin{eqnarray}
    \ket{\Psi} &=& \sum_{\vec{c}} a_{\vec{c}} \ket{\vec{c}}  \\
    &=&\sum_{\vec{c}} \frac{1}{\sqrt{\mu(\vec{c})}} a_{c_{1},\dots,c_{\Np}} \hat{a}_{c_{1}}^{\dagger} \dots \hat{a}_{c_{\Np}}^{\dagger} \ket{\bm{0}},
    \notag
\end{eqnarray}
where $\ket{\bm 0} = \ket{0}^{\otimes \Np}$ and
$\sum_{\vec{c}} (\ldots)\equiv \prod_{j=1}^{\Np}\sum_{c_{j}=0}^{M-1}(\ldots)$.
By applying the transformation \eqref{Scattering_trans}, each term $\ket{\vec{c}}$ in the superposition transforms as
\begin{eqnarray}
\label{scattered_c}
    && S(\ket{\vec{c}})   \\
    &&= \frac{1}{\sqrt{\mu \left(\vec{c} \right)}} \sum_{\vec{q}}  S_{q_{1}c_{1}} \dots S_{q_{\Np}c_{\Np}} \hat{a}_{q_{1}}^{\dagger} \dots \hat{a}_{q_{\Np}}^{\dagger} \ket{\bm 0} \, ,\notag
\end{eqnarray}
with $\mu(\vec{c}) = n_{1}!\cdot \dots \cdot n_{M}!$, where $n_{i}$ is the occupation number of the mode $i$ and $\sum_{i=0}^{M-1}n_{i}=\Np$.
Therefore, the transition amplitude to the state
\begin{equation}
\nonumber
\ket{ \vec{k} } = \frac{1}{\sqrt{\mu(\vec{k})}} \hat{a}^{\dagger}_{k_{1}}\dots \hat{a}^{\dagger}_{k_{\Np}} \ket{\bm 0}
\end{equation}
can be explicitly computed as $\braket{\vec{k}|S(\vec{c})}$. By using the bosonic canonical commutation relations
\begin{eqnarray}
    [\hat{a}_{i},\hat{a}_{j}^{\dagger}] = \delta_{ij},\quad [\hat{a}_{i},\hat{a}_{j}] = [\hat{a}_{i}^{\dagger},\hat{a}_{j}^{\dagger}] = 0
\end{eqnarray}
with some algebra we obtain
\begin{align}
\nonumber
     \braket{0|\hat{a}_{k_{1}} \dots \hat{a}_{k_{{\Np}}} \hat{a}_{q_{1}}^{\dagger}  \dots \hat{a}_{q_{{\Np}}}^{\dagger}|0} \notag \\
     \nonumber
    & \hspace{-1.5cm} =\displaystyle\sum_{ \{ q\} } \delta_{k_{1},q_{1}} \dots \delta_{k_{{\Np}},q_{{\Np}}}
\end{align}
where the sum runs over all the permutations of the indices $q_{1},\dots q_{\Np}$. Therefore, substituting the last expression in the Eq. \eqref{scattered_c}, the only terms that survive are the one with $\vec{q} = \vec{k}$ and all their permutations, which leads to
\begin{eqnarray}
    \braket{\vec{k}|S(\vec{c})} = \frac{1}{\sqrt{\mu\left(\vec{c} \right) \mu ( \vec{k} )}} \text{Perm}\left( S_{\vec{k}|\vec{c}}\right)
\end{eqnarray}
that is, Eq. \eqref{eq:perm_ampls}, and the transition probability is simply given by the squared modulus of the amplitude, leading to Eq. \eqref{eq:perm_probs}.


\subsection{Efficient probability calculation for fully-bunched model}
\label{FULLYBUNCHED}
Here we show how the probabilities can be efficiently calculated for the simplified model, shown in Fig. \ref{fig:lin_optics}d, where $\Lambda_{\rm FB}$ is composed of all on only the configuration where output photons are bunched together (i.e. are in the same output mode).

Because the input contains all photons in the same mode (assumed to be the $0$-th for simplicity), the probability for a generic output configuration $\vec{k}=\{k_1,\ldots,k_{\Np}\}$ can be calculated via Eq. (\ref{eq:perm_probs}) as:
\begin{align}
    \Pr(\vec{k}| { \bm \sigma}) &= \left| \bra{\vec{k}}\mathcal{U}(\ket{\vec{0}})  \right|^2 \\
    \nonumber
    &=  \frac{1}{ \Np! \mu(\vec{k})} \left|\text{Perm}\left( U_{\vec{k}|\vec{0}} \right) \right|^2 \\
    \nonumber
    & = \frac{\Np!}{ \mu(\vec{k})} 
    \prod_{i=1}^{\Np
    \left|S \cdot \text{diag}({\bm \sigma}) \cdot U^{\text{DFT}} \right|^2_{k_i, 0}} \\
    \nonumber
    &= \frac{\Np!}{M^\Np \mu(\vec{k})} 
      \prod_{i=1}^{\Np}
    \left|  S  {\bm \sigma} \right|^2_{k_i}
    \\
    \nonumber
    &=\frac{\Np!}{M^\Np \mu(\vec{k})}
     \prod_{i=1}^{\Np}
    \left| \sum_{j=0}^{M-1} S_{k_i,j}  \sigma_j \right|^2_{k_i},
\end{align}
where $\text{diag}(\bm \sigma)$ is a diagonal matrix having for entries the spin values $\sigma_1,\ldots, \sigma_M$
and
where we have used that for the Discrete Fourier Transform elements in the $0$-th column are simply given by $U^{\text{DFT}}_{i,0} = 1/\sqrt{M}$.  The situation is simplified even further when we consider an output configuration where all output photons bunch in the same $\mu$-th mode, i.e., in which $k_i=\mu\in[0,M-1]$, for each photon $i\in[1,\Np]$.

%
For this case the output probability is simplified to
\begin{equation}
    \Pr(\vec{k_\mu}|{ \bm \sigma}) = \frac{1}{M^\Np}
    \left| \sum_{j=0}^{M-1}  S_{\mu,j} \sigma_j \right|^{2\Np}.
    \label{eq:compiled_prob_app}
\end{equation}

Defining the set $\Lambda_{\rm FB} = \{\vec{k_0},\ldots,\vec{k_{M-1}}\}$ composed of all configurations with bunched photons on any of the modes, when considering the sum over more  output photonic configurations of this kind, composing a subset $\Lambda \subseteq \Lambda_{\rm FB}$  we obtain:
\begin{equation}
    \Pr(\Lambda|{ \bm \sigma}) = \frac{1}{M^\Np} \sum_{\mu\in [0,M-1]}
    \left| \sum_{j=0}^{M-1}  S_{\mu,j} \sigma_j \right|^{2\Np},
\end{equation}
where by $\mu\in[0,M-1]$ we mean that the index $\mu$ run through a subset of all possible modes in $[0,M-1]$.


\subsection{Time self-correlation function $F_{self}$}
\label{SELFCORR}
In order to analyze the spin correlations during time, we consider the time correlation function of spins, defined by
\begin{align}
  F(\tau) &= \sum_{i=1}^{M}\braket{\sigma_{i}(t)\sigma_{i}(t+\tau)}_{t} \notag \\
  &\equiv \frac{1}{N_{\rm MCs} \cdot N}\sum_{i=1}^{M}\sum_{t=0}^{N_{\rm MCs}}\sigma_{i}(t)\sigma_{i}(t+\tau), \quad
\end{align}
where $N_{\rm MCs}$ denotes the number of Monte Carlo steps, and the averaging is performed over the Monte Carlo steps $t$ and  $\tau\in [0,N_{\rm MCs}]$. The analysis has been carried out for different temperatures for both smaller and larger values of $\alpha$, with the results shown in Figs. \ref{fig:FselfError} a) and b), respectively.

 The correlation function decays rapidly to zero for high temperatures.
 At small $\alpha$, as the temperature decreases, the behavior exponential decay behavior changes into a short power-law decay to a plateau, as shown in Fig. \ref{fig:FselfError} a). That is, the system is stuck into one of the attractors and visits only configurations around that minimum.
For higher values of $\alpha$, Fig. \ref{fig:FselfError} b),
the speed at which the configuration decorrelates over time also slows down from exponential as the temperature decreases.  This is referred to as critical slowing down \cite{caltagirone2012critical,caltagirone2012bcritical,leuzzi2024logarithmic} in glassy physics and is a typical feature of many frustrated complex systems approaching a glassy phase.

\subsection{Retrieval order parameter $\hat{m}$}
\label{OVERLAP-M}

We define the parameter identifying a generalized {\em memory pattern} planted into the interferometric scattering system. Such a pattern is labeled by the output configuration $\vec k=\{k_1,k_2\}$ of the $2$-photon current and is the component $\vec k, \vec x$ of the permanent of the scattering matrix, that we shorten in the tensor
$$\mathbb X^{(\vec k)}_{\vec x}\equiv
\text{Perm} \left(S_{\vec{k}|\vec{x}}\right)=S_{k_1,x_1}S_{k_2,x_2} + S_{k_1,x_2}S_{k_2,x_1}.$$
The overlap of the input photonic mode phases $\bm\sigma=\{\sigma_1,\ldots, \sigma_N\}$ with such a matrix pattern is defined as

\begin{eqnarray}
    m_{\vec k} &\equiv& \frac{1}{M}\ \bm \sigma^T \mathbb X^{(\vec k)}~\bm \sigma
     \label{permanent-overlap}
     \\
    &=&
  \frac{1}{M}  \sum_{x_1,x_2}^{1,N}
     \sigma_{x_1}\left[
     S_{k_1,x_1}S_{k_2,x_2} + S_{k_1,x_2}S_{k_2,x_1}
     \right]
    \sigma_{x_2}
    \nonumber
\end{eqnarray}
In terms of (\ref{permanent-overlap}) the cost function, cf. Eq. (\ref{output_prob}), can be rewritten as
\begin{equation}
\mathcal H[{\bm\sigma}|\Lambda] =- M \mbox{Pr}(\Lambda|{\bm\sigma})
     =
M\sum_{\vec k \in \Lambda} \left| \hat m_{\vec k}\right|^2  ,
\label{hamiltonian-overlap}
\end{equation}
where the normalized overlaps
\begin{equation}
\hat m_{\vec k} \equiv   \frac{m_{\vec k} }{\sqrt{\sum_{k_1,k_2}^{1,M} |m_{\{k_1,k_2\}}|^2}}
\label{memory-overlap}
\end{equation}
  are introduced such that their real and imaginary parts are in the domain $[-1,1]$.

For every target $\vec k$ we will have a scattering matrix $S$, for every matrix $S$ a permanent $\mathbb X^{(\vec k)}_{\vec x}$, representing in the form of a $2$-index tensor (a $M \times M$ matrix of indices $x_1$ and $x_2$)
a memory planted into the neural network.
The Hamiltonian (\ref{hamiltonian-overlap}) is designed in such a way that its minima in ${\bm \sigma}$ are attracted to the maximum of a single $m_{\vec k}$, i.e., each $\text{Perm} \left(S_{\vec{k}| \ldots} \right)$ is an attractor of the ${\bm \sigma}$ dynamics in the energy landscape.

This is the case in the retrieval phase when such patterns are not too many. As $\alpha$ grows, though, new spurious attractors arise, not coinciding with the planted ones, and as the storage capacity of the network is saturated, they dominate the dynamics and become spin-glass states. Eventually, the original attractors are lost: this is the blackout phase.

\subsection{Spin-glass order parameter}
\label{OVERLAP-Q}
In the spin-glass phase, the minima $\bm \sigma$ of the Hamiltonian do not correspond to any scattering-matrix-permanent pattern. Spin-glass states have nothing to do with the scattering elements embedded into the Hopfield neural network by means of Eqs. (\ref{eq:J_ensemble},\ref{eq:J_ensemble_2_phot}).
 The memory overlaps (\ref{memory-overlap}) will all be vanishingly small, and there is no absolute reference to identify a state.
 For this reason, one resorts to the similarity between parallel dynamic histories of the system; that is, one ideally conceives the overlap between configurations evolving in exact {\em replicas} of the same system (same random matrix $S$). Such overlap order parameter does not refer to a state (an alignment, a vector pattern, a permanent of a matrix), but it is a similarity of two trajectories in the same corrugated energy landscape.

If we call $\bm \sigma^{(a)}$ the $M$ spin configuration of replica $a$,   the overlap is defined as
\begin{eqnarray}
\label{Qab}
    q_{ab}\equiv \frac{1}{M}\sum_{j=0}^{M-1} \sigma_j^{(a)} \sigma_j^{(b)} .
\end{eqnarray}
If replica symmetry holds, all states are equivalent and each thermalized couple of replicas will have the same overlap.
If the multi equilibria landscape is more complicated, as in structural glasses \cite{Parisi2020} and in proper spin-glasses \cite{Mezard1986,Mydosh1993}, replica symmetry is broken at low temperature and Eq. (\ref{Qab}) can take different values, with different probability,  to account for the complex organization of states. In this case the whole $P(q)$ distribution is the order parameter, else called the Parisi order parameter.

\subsection{Effects of noise}
\label{NOISE}
The measured probability  ${\rm Pr}(k_1,k_2)$, cf. Eq. (\ref{output_prob}), suffers of two different forms of uncertainty. The first one derives from the measurement uncertainty.
The error on the average frequency of realization of the photonic output configuration $\vec k$ can be simply estimated by repeating  $N_{\rm exp }$ random transmissions yielding or not yielding a counting of a configuration $\vec k$.
As a binomial process, the experimental average frequency of $\vec k$ realization will fluctuate around its theoretical average $\Pr(\vec k)$ with variance $\Pr(\vec k)(1-\Pr(\vec k))$. The experimental error on the average frequency of $\vec k$ will, therefore, be
\begin{align}
\label{marco-noise}
\sigma_{\rm exp}(\vec k)= \sqrt{\Pr(\vec k)(1-\Pr(\vec k))/N_{\rm exp}}.
\end{align}
If $\Pr$ is small, the process will tend to a Poissonian one, and this  will tend to
$\sigma_{\rm exp}(\vec k)  \simeq \sqrt{\Pr(\vec k)/N_{\rm exp}}$.

The second noise source comes from the dynamic evolution in the presence of a synaptic noise, represented by a temperature $T$ in the statistical physics modeling.
The temperature explicitly enters in the acceptance probability to overcome a barrier in the energy landscape, cf. Eq. (\ref{metropolis}). Reformulated in terms of the variation of the probability of a single output configuration $\vec{k}$ induced by a change of spin phases, $\Delta\mbox{Pr}(\vec k)=\mbox{Pr}(\vec k|\vec{\bm \sigma}(t))- \mbox{Pr}(\vec k|\vec{\bm \sigma}(t-1))$,   Eq.  (\ref{metropolis}) reads as
\begin{align}
\mbox{prob} = e^{\frac{ N{\Delta {\rm Pr}(\vec k)}}{T}}.
\end{align}
At fixed, non-zero temperature the long time values of  $\Pr(\vec k)$ will fluctuate around the average value at that temperature. The fluctuations being characterized by
$$ \sigma^2_T(\vec k) = \left\langle\left(\Delta {\rm Pr}(\vec k)\right)^2\right\rangle.$$

The typical probability depends on temperature, that is, on the noise on scattering channels or synaptic noise in the neural network mapping.
One can set a biunivocal correspondence between the average frequency $\Pr(\vec k)$ of a given $2$-photon configuration  and $T$ as the plateau value at long times at which the energy contribution of the scattering channels leading to $\vec k$, $\mathcal H[\bm \sigma|\vec k] = -M \Pr(\vec k|\bm \sigma)$, relaxes.
Therefore, a dependence $\sigma_{\rm exp}(T)$ can be established through Eq. (\ref{marco-noise}).

In experimental simulations, in which the two photon-current is employed to estimate ${\rm Pr}(\vec k)$, the user should take care that
\begin{equation}
 \sigma_{\rm exp}(\vec k) \ll \sigma_T(\vec k).
    \label{simulation-noise}
\end{equation}
As $N_{\rm exp}$ directly affects  $\sigma_{\rm exp}(\vec k)$, the experiment should be designed to encompass an ``Experimental time per iteration''  (ETI) so that $\sigma_{\rm exp}(\vec k)$ stays lower than $\sigma_T(\vec k)$.
\newpage

\subsection{Experimental realization proposal }
\label{EXPERIMENT}
\begin{figure}[ht!]
   \includegraphics[width=.9\columnwidth]{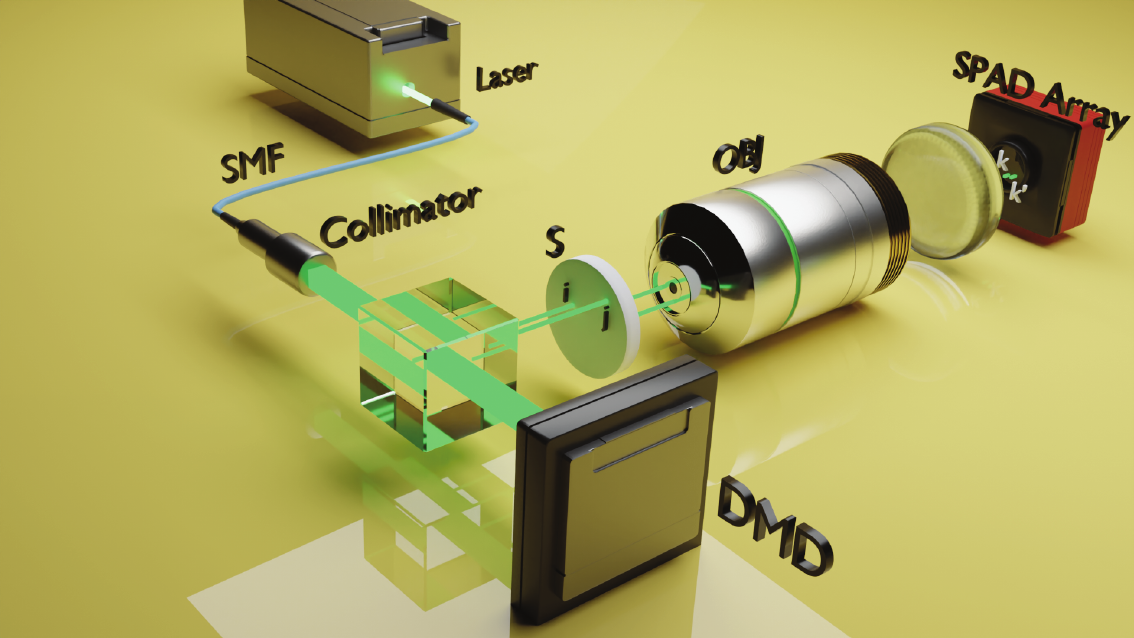}
    \caption{Sketch of the experimental setup.  SMF: single mode fiber;  DMD : Digital Micromirror Device; S: Sample; OBJ: Objective; $i$ and $j$ represent the populated mode index.} 
    \label{fig: Sketch}
\end{figure}

A proposal for the experimental realization of the experiment is depicted in the sketch in Fig. \ref{fig: Sketch}.  Laser light is delivered to our quantum simulator through a single-mode optical fiber, where it is collimated through an appropriate lens system, thus generating a homogeneous superposition spatial state. Because the input photons are all considered to be injected in the same input mode, their generation can be conveniently achieved by injecting a weak-coherent state into that mode, obtained by strong attenuation of a laser, and then post-selecting on the photon pairs component of it through the output photo-detection. The state is modulated through a spatial light modulator or a Digital micromirror device in phase modulation configuration, acting on light modes, adding $\pi$ or 0 phase delay. The modulator is imaged onto an optically disordered medium (white paint opaque layer) or a multi-mode optical fiber. The output of the scattering system is imaged through an objective to the Detector. The two-photon current can be detected at the output modes $[k, k']$, e.g., by a SPAD array.  The computational bottlenecks for this quantum simulator are the DMD Speed, the SPAD array count rate, and the electronics correlation speed.

\subsection{Other sources of non-ideality}
Losses from imperfect sources, dark counts, and chip coupling, as well as propagation losses, all contribute to a decrease in count rate but primarily affect the overall integration time rather than the algorithm's scaling \cite{weninger2025low,oh2025recent,wang2019boson,alexander2404manufacturable,gines2022high}. While the purity and indistinguishability of single photons \cite{zhang2018design} can be very high, residual distinguishability and photon loss generally do not impact the protocol's performance with an increasing number of modes (current technology provides, at our knowledge the possibility to reach 216 modes in a single chip \cite{madsen2022quantum}), only becoming significant when the number of photons increases. Current technology allows for precise phase shifter accuracy, leading to nearly error-free spin state control. Regarding the achievable system speed, a major bottleneck for photonic computing is due to the slow pace of heating based phase control. Employing the micro electro mechanical or lithium niobate technology optical elements for the spin related phase control system speed can be boosted by a factor $10^5$ to $>10^9$ \cite{zhang2021integrated}.

\begin{figure}
\includegraphics[width = 0.9\columnwidth]{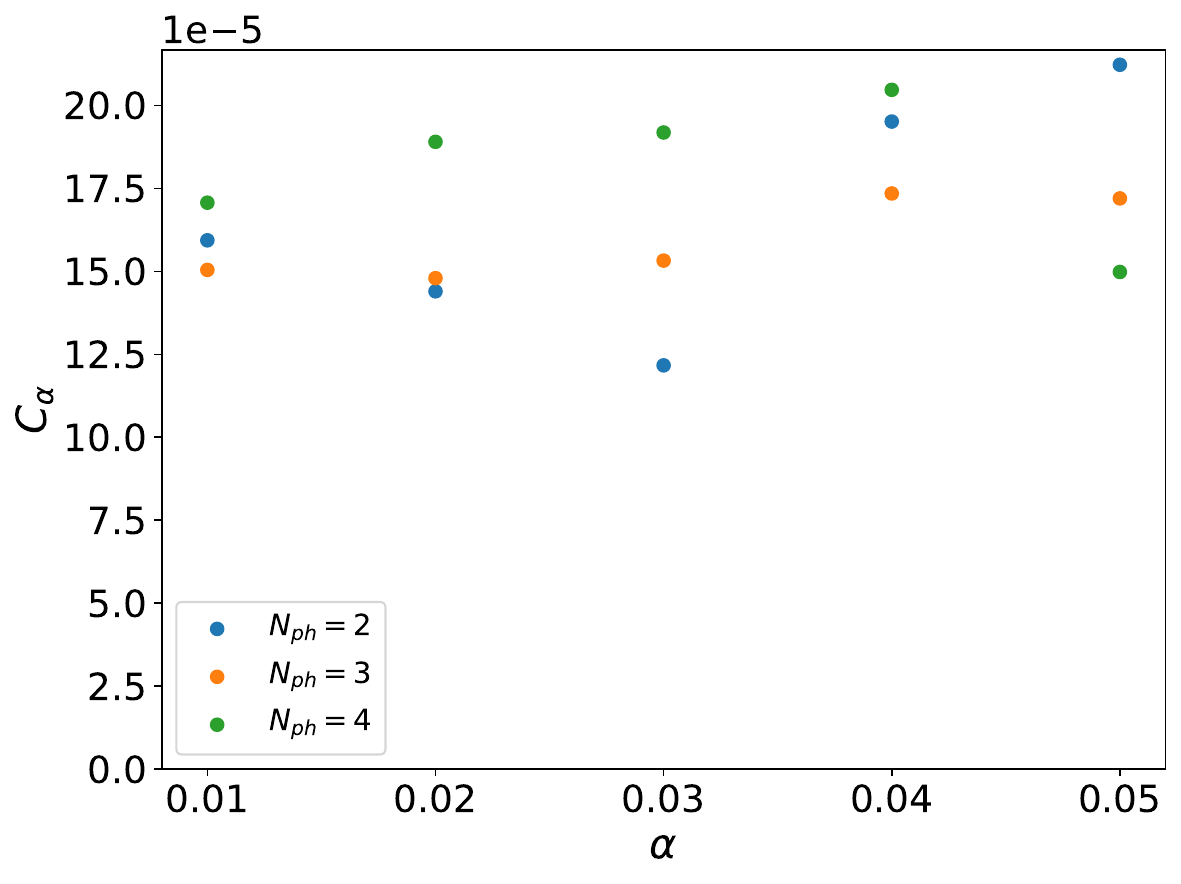}
\caption{ Values of the coefficient $C_\alpha\simeq t_{\rm MCU}/|\Lambda |$ of the scaling with $\alpha M^\Np$ of the computational cost of Monte Carlo Update in the digital simulation of the dynamics.}
\label{fig:c_alpha}
\end{figure}

\subsection{Scaling Performance}
The calculation on a digital computer of the energy of a many body Hopfield model, requires a large number of floating point operations.

Once a spin $\sigma_i$ is updated
the new energy is 
\begin{eqnarray}
\label{output_prob_H}
 \Pr(\Lambda |{\bm\sigma} ) &=& \sum_{\vec{k}\in \Lambda}\Pr_{\mathcal S} (\vec{k} |{\bm\sigma} )
= \sum_{\vec{k}\in \Lambda}
    \left|\bra{\vec{k}} \mathcal{S}(\ket{\psi'})\right|^2  
    \\
    \nonumber
    &=&
 \sum_{\vec{k}\in \Lambda}   \sum_{\vec{x},\vec{y}\in C_M}  \frac{ a_{\vec{x}}  a^*_{\vec{y}}}{\sqrt{\mu(\vec{x}) \mu(\vec{y})\mu(\vec{k})^2}} 
    \\
    \nonumber
    && \times \text{Perm} \left(S_{\vec{k}|\vec{x}}\right)
    \text{Perm} \left(S^*_{\vec{k}|\vec{y}}\right)\prod_{j=1}^{\Np}\sigma_{x_j}\sigma_{y_j}
    \\
    \nonumber
    &\simeq&
\frac{\Np !}{M^{\Np}} 
\sum_{\vec{k}\in \Lambda}    \sum_{\vec{x},\vec{y}\in C_M}  \frac{ 1}{\mu(\vec{x}) \mu(\vec{y})\mu(\vec{k})}
 \\
    \nonumber
    && \times \text{Perm} \left(S_{\vec{k}|\vec{x}}\right)
    \text{Perm} \left(S^*_{\vec{k}|\vec{y}}\right)\prod_{j=1}^{\Np}\sigma_{x_j}\sigma_{y_j}
\end{eqnarray}
\noindent 
in which only spin $\sigma_i$ in the $M$-mode configuration $\bm \sigma $ has changed sign.

Profiling our most efficient code for the (digital) simulation of the Monte Carl dynamics in the Hamiltonian landscape $\mathcal H[\bm \sigma]\equiv - M \Pr(\Lambda |{\bm\sigma} ) $ the time it takes to perform a single spin update of the Metropolis algorithm (Monte Carlo Update), including the computation of Eq. (\ref{output_prob_H}), turns out to scale as 
\begin{align}
t_{MCU} \simeq C_\alpha \, \alpha\,  M^{\Np} \simeq  C_\alpha \, |\Lambda|, 
\end{align}
with the number of modes $M$, the number of interfering photons $\Np$ and the relative capacity $\alpha\equiv |\Lambda|M^{-\Np}$. Else said, the computational cost grows
 proportionally to the number $P=|\Lambda|$ of acquired outputs. As shown in Fig. \ref{fig:c_alpha} the proportionality coefficient $C_\alpha$ does not strongly depend on $\Np$, at least  in the range of simulated $M$'s and $\Np$'s.

It is important to emphasize, however, that \sout{this} the increase in computational cost does not directly carry over to our experimental quantum simulator. In the analog photonic system, raising $\Np$ does not extend the measurement time, which remains limited by the same bottlenecks as in the $\Np=2$ case. Although using more photons increases the experimental complexity, the key advantage is that the performance of our photonic computer scales exponentially with the number of modes.

\begin{figure}[h]
\includegraphics[width=.9\columnwidth]{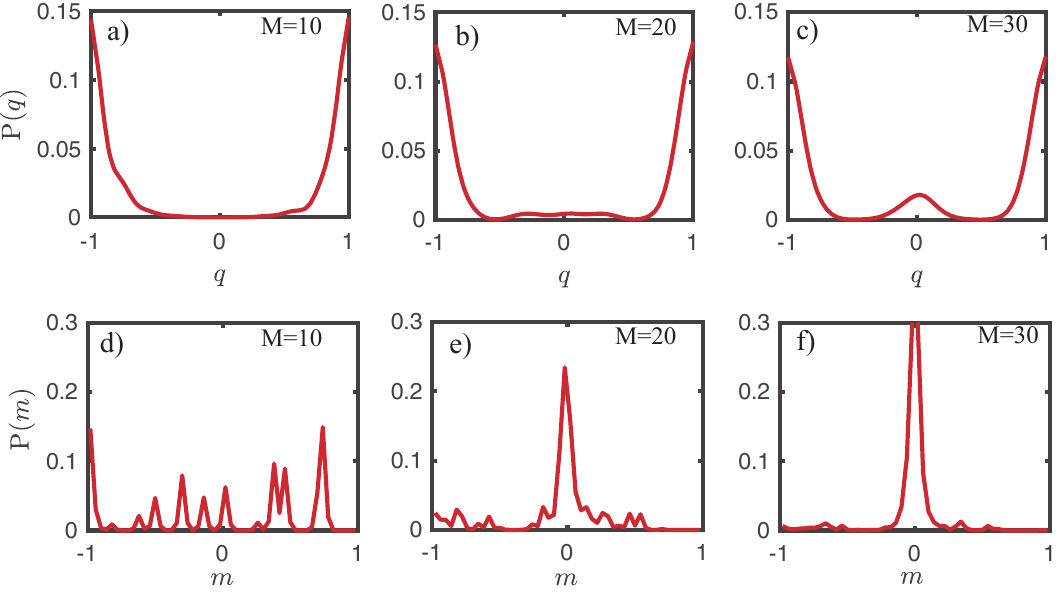}
\caption{ $P(q)$ and $P(m)$ for various values of the number of modes $M$, in the $\alpha=0.01$ and $T=0.075$ point of the phase diagram. }
\label{fig: Finite size effects}
\end{figure}

\subsection{Finite Size Effects}
 Many predictions for Hopfield and Ising models, are derived within the mean-field approximation, which assumes an infinite number of elements (M $\rightarrow \infty$). Nevertheless  A thorough characterization of finite-size effects in our system would require extensive experiments, we report here evidence of the the most prominent effect:the Retrieval–Spin Glass transition towards larger values of $\alpha$ as M decreases. This effect is reported in Fig. \ref{fig: Finite size effects}  in which we show $P(q)$ and $P(m)$ for $\alpha=0.01$ and $T=0.075$. In the phase diagram reported in figure 3 of the main paper. this point steadily resides in the Spin Glass phase. However when M decreases it is possible to note that $P(q)$ and  $P(m)$ denote the Retrieval regime.

\end{document}